\documentclass[10pt, final, conference, letterpaper, onecolumn, oneside]{./IEEEtran}

\usepackage{cite}
\usepackage{graphicx}
\usepackage{subfigure}
\usepackage{amsmath}
\usepackage{amssymb}
\usepackage{bm}
\usepackage{epsfig}
\usepackage{times}
\usepackage{color}
\usepackage{url}
\usepackage{array}
\usepackage{multirow}
\usepackage{rotating}

\newtheorem{pavikc}{\textbf{Corollary}}
\newtheorem{pavikl}{\textbf{Lemma}}

\newtheorem{pavikp}{\textbf{Proposition}}

\newcommand{\argmax}{\operatornamewithlimits{argmax}}
\newcommand{\pd}[2]{\frac{\partial#1}{\partial#2}}

\begin{document}

\title{Analog Network Coding in General SNR Regime}%
\author{\IEEEauthorblockN{Samar Agnihotri, Sidharth Jaggi, and Minghua Chen}\\%
\IEEEauthorblockA{Department of Information Engineering, The Chinese University of Hong Kong, Hong Kong\\}%
Email: samar.agnihotri@gmail.com, \{jaggi, minghua\}@ie.cuhk.edu.hk%
}

\maketitle

\begin{abstract}
The problem of maximum rate achievable with analog network coding for a unicast communication over a layered wireless relay network with directed links is considered. A relay node performing analog network coding scales and forwards the signals received at its input. Recently this problem has been considered under two assumptions: (A) each relay node scales its received signal to the upper bound of its transmit power constraint, (B) the relay nodes in specific subsets of the network operate in the high-SNR regime. We establish that assumption (A), in general, leads to suboptimal end-to-end rate. We also characterize the performance of analog network coding in class of symmetric layered networks without assumption (B).

The key contribution of this work is a lemma that states that a globally optimal set of scaling factors for the nodes in a layered relay network that maximizes the end-to-end rate can be computed layer-by-layer. Specifically, a rate-optimal set of scaling factors for the nodes in a layer is the one that maximizes the \textit{sum-rate} of the nodes in the next layer. This critical insight allows us to characterize analog network coding performance in network scenarios beyond those that can be analyzed using the existing approaches. We illustrate this by computing the maximum rate achievable with analog network coding in one particular layered network, in various communication scenarios.
\end{abstract}

\section{Introduction}
\label{sec:intro}
Analog network coding (ANC) extends to multihop wireless networks the idea of linear network coding \cite{103liYeungCai} where an intermediate node sends out a linear combination of its incoming packets. In a wireless network, signals transmitted simultaneously by multiple sources add in the air. Each node receives a \textit{noisy sum} of these signals, \textit{i.e.} a linear combination of the received signals and noise. A multihop relay scheme where an intermediate relay node merely amplifies and forwards this noisy sum is referred to as analog network coding \cite{107kattiGollakottaKatabi, 110maricGoldsmithMedard}.

The performance of the analog network coding in layered relay networks is previously analyzed in \cite{110maricGoldsmithMedard, 111liuCai}. In \cite{110maricGoldsmithMedard}, the achievable rate is computed under two assumptions: (A) each relay node scales the received signal to the maximum extent possible subject to its transmit power constraint, (B) the nodes in all layers operate in the high-SNR regime, where the received signal power $P_{R,k}$ at the $k^\textrm{th}$ node satisfies $\min_{k \in l} P_{R,k} \ge 1/\delta, l = 1, \ldots, L$ for some small $\delta \ge 0$, where $L$ is the number of layers of relay nodes. It is shown that the rate achieved under these two assumptions approaches network capacity as the source power increases. The authors in \cite{111liuCai} extend this work to the scenarios where the nodes in at most one layer do not satisfy these two assumptions and show that achievable rates in such scenarios also approach the network capacity as the source power increases.

However, requiring each relay node to amplify its received signal to the upper bound of its transmit power constraint results, in general, in suboptimal end-to-end performance of analog network coding, as we establish in this paper and also previously indicated in \cite{101schein, 111agnihotriJaggiChen}. Further, even in low-SNR regimes amplify-and-forward relaying can be capacity achieving relay strategy in some scenarios, \cite{107gomadamJafar}. Therefore, in this paper we are concerned with analyzing the performance of analog network coding in layered networks, without above two assumptions on input signal scaling factors and received SNRs. Computing the maximum rate achievable with analog network coding without these two assumptions, however, results in a computationally intractable problem, in general \cite{111liuCai, 111agnihotriJaggiChen}.

Our main contribution is a result that states that a globally optimal set of scaling factors for the nodes that maximizes the end-to-end rate in a general layered relay network can be computed layer-by-layer. In particular, a rate-optimal set of scaling factors for the nodes in a layer is the one that maximizes the sum-rate of the nodes in the next layer. This result allows us to exactly compute the optimal end-to-end rate achievable with analog network coding, over all possible choices of scaling factors for the nodes, in a class of layered networks that cannot be so addressed using existing approaches. We illustrate this by computing the maximum ANC rate in different scenarios for one particular layered network. Further, for general layered relay networks, our result significantly reduces the computational complexity of solving this problem.

\textit{Organization:} In Section~\ref{sec:sysModel} we introduce a general wireless layered relay network model and formulate the problem of maximum rate achievable with ANC in such a network. We also provide an example to illustrate the complete problem formulation for a specific layered relay network. Section~\ref{sec:computingANCperf} discusses the computational hardness of this problem and existing approaches to address it. In Section~\ref{sec:betaComputation} we first motivate and then state and prove the key lemma of this paper that allows us to compute a rate-optimal set of scaling factors for the nodes in a layered network in a layer-by-layer manner. Then Section~\ref{sec:exa} illustrates the computation of the maximum ANC rate in one particular layered network in various scenarios. Finally, Section~\ref{sec:conclFW} concludes the paper.

\section{System Model}
\label{sec:sysModel}
\begin{figure}[!t]
\centering
\includegraphics[width=3.0in]{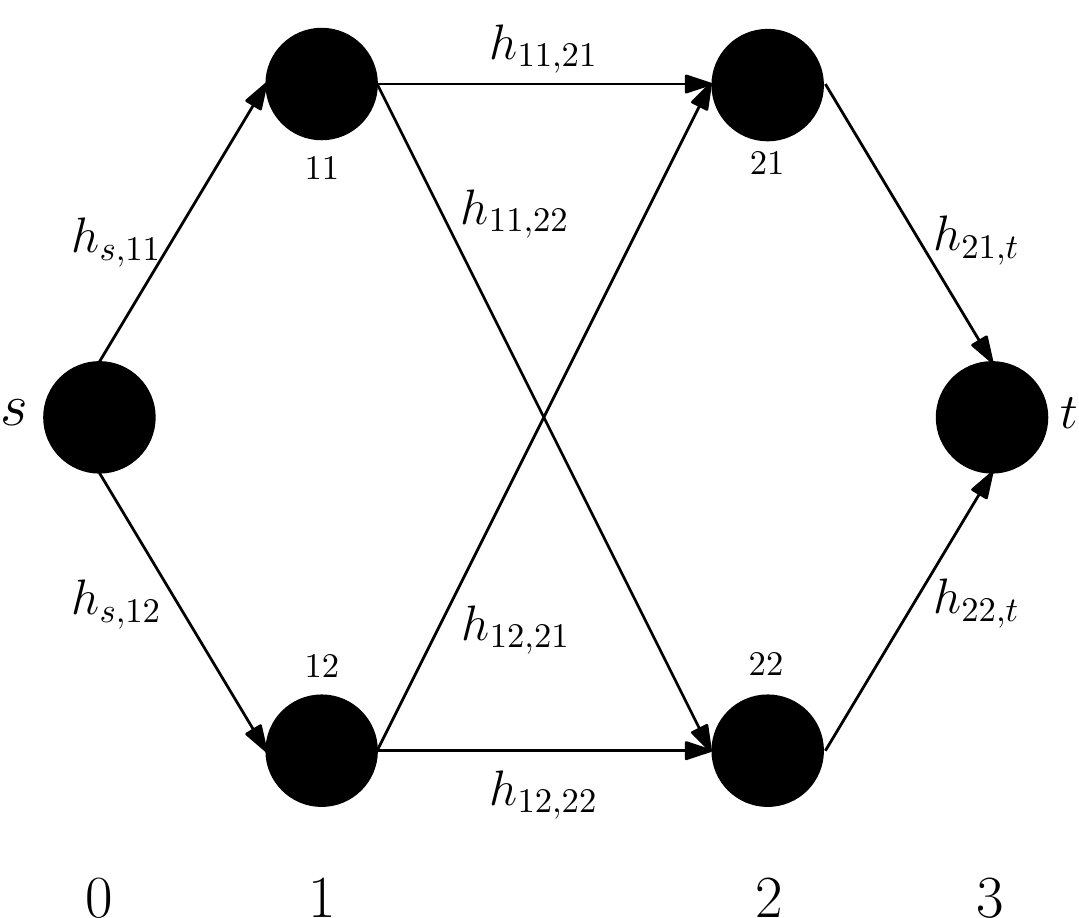}
\caption{Layered network with $L=2$ layers of relay nodes between source `s' and destination `t'. Each layer contains two relay nodes. Source is at layer `0' and destination at layer `3'. The $i^\textrm{th}$ node in the $l^\textrm{th}$ layer is denoted as $li$, $l \in \{1,2\}, i \in \{1,2\}$}.
\label{fig:sdChannelExa}
\end{figure}

Consider a $(L+2)$-layer wireless relay network with directed links\footnote{The layered networks with bidirected links can be addressed with the \textit{signal subtraction} notion we introduced in \cite{111agnihotriJaggiChen}. However, for the ease of presentation we do not discuss such networks in this paper.}. Source $s$ is at layer `$0$', destination $t$ is at layer `$L+1$', and the relay nodes from the set $R$ are arranged in $L$ layers between them. The $l^\textrm{th}$ layer contains $n_l$ relay nodes, $\sum_{l=1}^L n_l = |R|$. An instance of such a network is given in Figure~\ref{fig:sdChannelExa}. Each node is assumed to have a single antenna and operate in full-duplex mode.

At instant $n$, the channel output at node $i, i \in R \cup \{t\}$, is
\begin{equation}
\label{eqn:channelOut}
y_i[n] = \sum_{j \in {\mathcal N}(i)} h_{ji} x_j[n] + z_i[n], \quad - \infty < n < \infty,
\end{equation}
where $x_j[n]$ is the channel input of node $j$ in neighbor set ${\mathcal N}(i)$ of node $i$. In \eqref{eqn:channelOut}, $h_{ji}$ is a real number representing the channel gain along the link from node $j$ to node $i$. It is assumed to be fixed (for example, as in a single realization of a fading process) and known throughout the network. The source symbols $x_s[n], - \infty < n < \infty$, are independently and identically distributed (\textit{i.i.d.}) Gaussian random variables with zero mean and variance $P_s$ that satisfy an average source power constraint, $x_s[n] \sim {\cal N}(0, P_s)$. Further, $\{z_i[n]\}$ is a sequence (in $n$) of \textit{i.i.d.} Gaussian random variables with zero mean and variance $\sigma^2, z_i[n] \sim {\cal N}(0, \sigma^2)$. We also assume that $z_i$ are independent of the input signal and of each other. We assume that the $i^{\textrm{th}}$ relay's transmit power is constrained as:
\begin{equation}
\label{eqn:pwrConstraint}
E[x_i^2[n]] \le P_i, \quad - \infty < n < \infty
\end{equation}

In analog network coding, each relay node amplifies and forwards the noisy signal sum received at its input. More precisely, a relay node $i$ at instant $n+1$ transmits the scaled version of $y_i[n]$, its input at time instant $n$, as follows
\begin{equation}
\label{eqn:AFdef}
x_i[n+1] = \beta_i y_i[n], \quad 0 \le \beta_i^2 \le \beta_{i,max}^2 = P_i/P_{R,i},
\end{equation}
where $P_{R,i}$ is the received power at node $i$ and the scaling factor $\beta_i$ is chosen such that the power constraint \eqref{eqn:pwrConstraint} is satisfied.

One important characteristic of layered networks is that all paths from the source to the destination have the same number of hops. Also, each path from the $i^\textrm{th}, i \in R$, relay node to the destination has the same length. In other words, in a layered network with $L$ layers, all copies of a source signal traveling along different paths arrive at the destination with time delay $L$ and all copies of a noise symbol introduced at a node in $l^\textrm{th}$ layer arrive at the destination with time delay $L-i+1$. Therefore, all outputs of the source-destination channel are free of intersymbol interference. This simplifies the relation between input and output of the source-destination channel and allows us to omit the time-index while denoting the input and output signals.

Using \eqref{eqn:channelOut} and \eqref{eqn:AFdef}, the input-output channel between the source and the destination can be written as
\begin{equation}
\label{eqn:sdchnl}
y_t = \bigg[\sum_{(i_1, \ldots, i_{L}) \in K_{s}} \hspace{-0.15in} h_{s,i_1}\beta_{i_1}h_{i_1,i_2} \ldots \beta_{i_{L}}h_{i_{L},t}\bigg] x_s  + \sum_{l=1}^{L} \sum_{j=1}^{n_l}\bigg[\sum_{(i_1, \ldots, i_{L-l+1}) \in K_{lj}} \hspace{-0.35in} \beta_{i_1} h_{i_1, i_2} \ldots \beta_{i_{L-l+1}} h_{i_{L-l+1}, t}\bigg] z_{lj} + z_t, 
\end{equation}
where $K_s$ is the set of $L$-tuples of node indices corresponding to all paths from source $s$ to destination $t$ with path delay $L$. Similarly, $K_{lj}$ is the set of $L-l+1$-tuples of node indices corresponding to all paths from the $j^{\textrm{th}}$ relay of $l^\textrm{th}$ layer to destination $t$ with path delay $L-l+1$.

We introduce \textit{modified} channel gains as follows. For all the paths between source $s$ and destination $t$:
\begin{equation}
\label{eqn:modChnlParams}
h_s = \sum_{(i_1, \ldots, i_{L}) \in K_{s}} h_{s,i_1}\beta_{i_1}h_{i_1,i_2} \ldots \beta_{i_{L}}h_{i_{L},t}
\end{equation}
For all the paths between the $j^{\textrm{th}}$ relay of $l^\textrm{th}$ layer to destination $t$ with path delay $L-l+1$:
\begin{equation}
\label{eqn:modChnlParams2}
h_{lj} = \sum_{(i_1, \ldots, i_{L-l+1}) \in K_{lj}} \beta_{i_1} h_{i_1, i_2} \ldots \beta_{i_{L-l+1}} h_{i_{L-l+1}, t}
\end{equation}

In terms of these modified channel gains\footnote{Modified channel gains for even a possibly exponential number of paths as in \eqref{eqn:modChnlParams} and \eqref{eqn:modChnlParams2} can be efficiently computed using the line-graphs \cite{103koetterMedard}, and there are only a polynomial number of them in polynomial sized graph.}, the source-destination channel in \eqref{eqn:sdchnl} can be written as:
\begin{equation}
\label{eqn:chnlmod}
y_t = h_s x_s + \sum_{l=1}^{L} \sum_{j=1}^{n_l} h_{lj} z_{lj} + z_t 
\end{equation}

\textit{Problem Formulation:} For a given network-wide scaling vector $\bm{\beta}=(\beta_{lj})_{1 \le l \le L, 1 \le j \le n_l}$, the achievable rate for the channel in \eqref{eqn:chnlmod} with \textit{i.i.d.} Gaussian input is (\hspace{-0.001cm}\cite{110maricGoldsmithMedard, 111liuCai, 111agnihotriJaggiChen}):
\begin{equation}
\label{eqn:infoRateFin}
I(P_s, \bm{\beta}) = \frac{1}{2} \log\big(1 + SNR_t\big),
\end{equation}
where $SNR_t$, the signal-to-noise ratio at destination $t$ is given by
\begin{equation}
\label{eqn:snr}
SNR_t = \frac{P_s}{\sigma^2} \frac{h_s^2}{1 + \sum_{l=1}^{L} \sum_{j=1}^{n_l} h_{lj}^2}
\end{equation}

For a given network-wide scaling vector $\bm{\beta}$, the achievable rate of information transfer is given by $I(P_s, \bm{\beta})$. Therefore the maximum information-rate $I_{ANC}(P_s)$ achievable with analog network coding in a given layered network with \textit{i.i.d.} Gaussian input is defined as the maximum of $I(P_s, \bm{\beta})$ over all feasible $\bm{\beta}$, subject to per relay transmit power constraint \eqref{eqn:AFdef}. In other words:
\begin{equation}
\label{eqn:maxAFrate}
\mbox{(P1): } \qquad\qquad I_{ANC}(P_s) \stackrel{def}{=} \max_{\bm{\beta}:0 \le \beta_{lj}^2 \le \beta_{lj, max}^2} I(P_s, \bm{\beta})
\end{equation}
It should be noted that $\beta_{lj, max}^2$ (the maximum value of the scaling factor for $j^\textrm{th}$ node in the $l^\textrm{th}$ layer) depends on the scaling factors of the nodes in the previous $l-1$ layers as illustrated in the example below.

Next, we provide an example to illustrate the derivation of the source-destination channel expression in \eqref{eqn:chnlmod} and the problem formulation \eqref{eqn:maxAFrate} for a specific layered network.

\textbf{\textit{Example 1:}} Consider the layered network in Figure~\ref{fig:sdChannelExa}. For this network we have:
\begin{align*}
K_s &= \{(11,21), (11,22), (12,21), (12,22)\}, \\
K_{11} &= \{(11,21), (11,22))\}, \\
K_{12} &= \{(12,21), (12,22)\}, \\
K_{21} &= \{(21)\}, \\
K_{22} &= \{(22)\}
\end{align*}
Then, we have the following expressions for the modified channels gains from the source and each relay node to the destination
\begin{align*}
h_s &= h_{s,11} \beta_{11} h_{11,21} \beta_{21} h_{21,t} + h_{s,11} \beta_{11} h_{11,22} \beta_{22} h_{22,t} + h_{s,12} \beta_{12} h_{12,21} \beta_{21} h_{21,t} + h_{s,12} \beta_{12} h_{12,22} \beta_{22} h_{22,t}, \\
h_{11} &= \beta_{11} h_{11,21} \beta_{21} h_{21,t} + \beta_{11} h_{11,22} \beta_{22} h_{22,t}, \\
h_{12} &= \beta_{12} h_{12,21} \beta_{21} h_{21,t} + \beta_{12} h_{12,22} \beta_{22} h_{22,t}, \\
h_{21} &= \beta_{21} h_{21,t}, \\
h_{22} &= \beta_{22} h_{22,t}
\end{align*}
Thus, we have
\begin{equation*}
y_t = h_s x_s + \sum_{l=1}^{2} \sum_{j=1}^{2} h_{lj} z_{lj} + z_t 
\end{equation*}
as an expression for the source-destination channel in this case.

The maximum information-rate $I_{ANC}(P_s)$ achievable with analog network coding in the layered network of Figure~\ref{fig:sdChannelExa} with \textit{i.i.d.} Gaussian input is the solution of the following problem
\begin{equation*}
I_{ANC}(P_s) \stackrel{def}{=} \max_{\bm{\beta}:0 \le \beta_{lj}^2 \le \beta_{lj, max}^2} \frac{1}{2} \log\big(1 + SNR_t\big),
\end{equation*}
where
\begin{equation*}
SNR_t = \frac{P_s}{\sigma^2} \frac{h_s^2}{1 + \sum_{l=1}^{2} \sum_{j=1}^{2} h_{lj}^2}
\end{equation*}
is the SNR at destination $t$ and
\begin{align*}
\beta_{11, max}^2 &= \frac{P_{11}}{h_{s,11}^2 P_s + \sigma^2} \\
\beta_{12, max}^2 &= \frac{P_{12}}{h_{s,12}^2 P_s + \sigma^2} \\
\beta_{21, max}^2 &= \frac{P_{21}}{(h_{s,11}^2 \beta_{11}^2 h_{11,21}^2 + h_{s,12}^2 \beta_{12}^2 h_{12,21}^2)P_s + (\beta_{11}^2 h_{11,21}^2 + \beta_{12}^2 h_{12,21}^2 + 1) \sigma^2} \\
\beta_{22, max}^2 &= \frac{P_{22}}{(h_{s,11}^2 \beta_{11}^2 h_{11,22}^2 + h_{s,12}^2 \beta_{12}^2 h_{12,22}^2)P_s + (\beta_{11}^2 h_{11,22}^2 + \beta_{12}^2 h_{12,22}^2 + 1) \sigma^2}
\end{align*}
are the maximum values of the scaling-factors for the relay nodes. {\hspace*{\fill}~\IEEEQEDclosed\par}

Given the monotonicity of the $\log(\cdot)$ function, we have
\begin{equation}
\label{eqn:eqProb}
\mbox{(P2): } \qquad\qquad \bm{\beta}_{opt} = \argmax_{\bm{\beta}:0 \le \beta_{lj}^2 \le \beta_{lj, max}^2}  I(P_s, \bm{\beta}) = \argmax_{\bm{\beta}:0 \le \beta_{lj}^2 \le \beta_{lj, max}^2}  SNR_t
\end{equation}
Therefore in the rest of the paper, we concern ourselves mostly with maximizing the received SNRs.

\section{Analyzing the optimal performance of analog network coding in general layered networks}
\label{sec:computingANCperf}
The problem \eqref{eqn:eqProb} is a hard optimization problem. In terms of \textit{Geometric Programming} \cite{105chiang, 107boydkimVandenberghe}, $SNR_t$ is a ratio of \textit{posynomials} that is a nonlinear (neither convex nor concave) function of $\sum_l n_l$ variables in $\bm{\beta}$, in general. It is well-known that maximizing such ratios of posynomials is an intractable problem with no efficient and global solution methods \cite[Page 85]{105chiang}. However, globally optimal solutions of such problems can be approximated using heuristic methods based on \textit{signomial programming} condensation that solves a sequence of geometric programs, as in \cite[Section 3.3]{105chiang}. Such heuristics though useful in providing \textit{good} numerical approximations to the optimal achievable $SNR_t$, do not provide non-trivial characterization of the optimal $SNR_t$ (or an optimal $\bm{\beta}_{opt}$ that achieves it) in terms of various system parameters. We argue that such characterization however, is highly desired not only for the accurate analysis of ANC performance in general layered networks, but also for the following reasons:
\begin{itemize}
\item non-trivial characterization of the performance of analog network coding in general wireless relay networks (non-layered with bidirectional links) in general SNR regime.
\item providing insights about the optimal relay operation, thus helping in the design of optimal relay schemes (\cite{107gomadamJafar, 109cuiHoKliewer}).
\item construction of distributed schemes to compute $\bm{\beta}_{opt}$.
\end{itemize}

Towards this goal, in \cite{110maricGoldsmithMedard, 111liuCai} the performance of analog network coding is analyzed under assumptions A and B discussed earlier about per node scaling factor and received SNR at each node, respectively.

In the following, we first provide an example to establish that assumption A, in general, leads to suboptimal ANC rates. Then, in the next section we introduce our result that allows us to analyze the optimal performance of analog network coding in a wide class of layered networks without assumption B or its limited relaxation in \cite{111liuCai}. This result also provides some key insights into the nature of $\bm{\beta}_{opt}$ in terms of various system parameters, allowing us to make progress towards addressing some of the objectives mentioned above.

\begin{figure}[!t]
\centering
\includegraphics[width=2.0in]{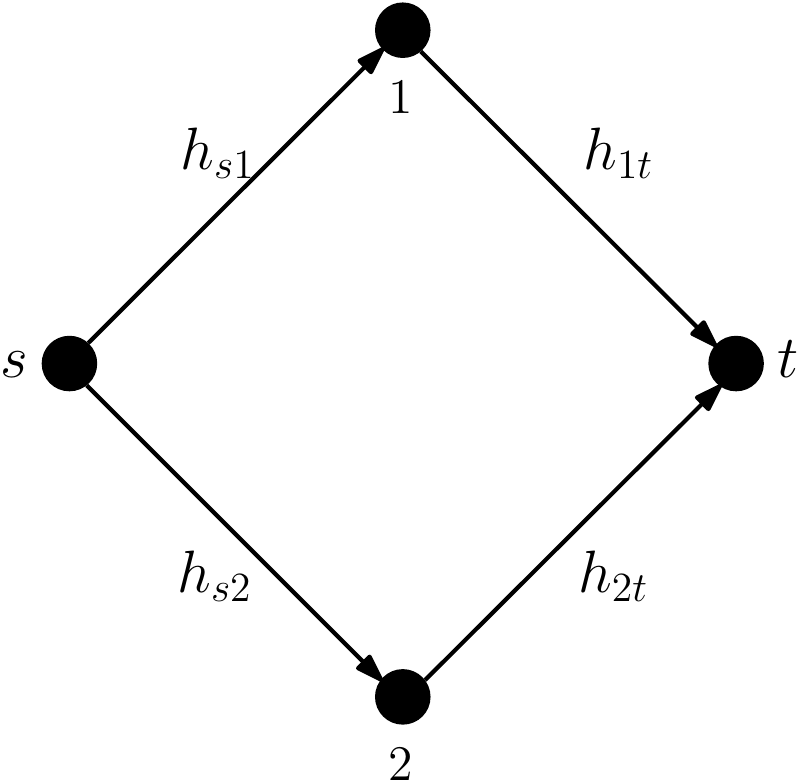}
\caption{A 2-relay Gaussian diamond network.}
\label{fig:diamond}
\end{figure}

\textbf{\textit{Example 2:}} Consider the 2-relay Gaussian diamond network, \cite{110maricGoldsmithMedard, 101schein}, in Figure~\ref{fig:diamond}. It is defined as a directed graph $G = (V,E)$ with $V = \{s, t, 1, 2\}$ and $E=\{(s,1), (s,2), (1,t), (2,t)\}$. Let $h_{e}$ be the channel gain along the link $e, e \in E$. The SNR at destination $t$ for this network is given as
\begin{equation*}
SNR_t = \frac{P_s}{\sigma^2} \frac{(h_{s1} \beta_1 h_{1t} + h_{s2} \beta_2 h_{2t})^2}{1 + \beta_1^2 h_{1t}^2 + \beta_2^2 h_{2t}^2}
\end{equation*}
Therefore, using \eqref{eqn:eqProb} the problem of maximum rate achievable with analog network coding for this network can be formulated as
\begin{equation}
\label{eqn:diamondProb}
\argmax_{0 \le \bm{\beta}^2 \le \bm{\beta}_{max}^2} SNR_t,
\end{equation}
where $\bm{\beta} = (\beta_1, \beta_2)$ and $\bm{\beta}_{max} = (\beta_{1,max}, \beta_{2,max})$ with $\beta_{1,max}^2 = P_1/(h_{s1}^2 P + \sigma^2), \beta_{2,max}^2 = P_2/(h_{s2}^2 P + \sigma^2)$.

Equating the first-order partial derivatives of the objective function with respect to $\beta_1$ and $\beta_2$ to zero, we get the following three conditions for global extrema:
\begin{eqnarray}
\beta_1 &=& - \frac{h_{s2} h_{2t}}{h_{s1} h_{1t}} \beta_2 \label{eqn:b1b2} \\
\beta_1 &=& \frac{h_{s1}}{h_{s2} h_{1t} h_{2t} \beta_2} + \frac{h_{s1} h_{2t}}{h_{s2} h_{1t}} \beta_2 \label{eqn:b1itob2} \\
\beta_2 &=& \frac{h_{s2}}{h_{s1} h_{1t} h_{2t} \beta_1} + \frac{h_{s2} h_{1t}}{h_{s1} h_{2t}} \beta_1 \label{eqn:b2itob1}
\end{eqnarray}

%
%

Denote the second-order partial derivatives of the objective function with respect to $\beta_1$ and $\beta_2$ as follows:
\begin{eqnarray*}
SNR_{\beta_1 \beta_1} = \frac{\partial^2 SNR_t}{\partial^2 \beta_1}, \quad SNR_{\beta_1 \beta_2} = \frac{\partial^2 SNR_t}{\partial \beta_1 \partial \beta_2} \\
SNR_{\beta_2 \beta_1} = \frac{\partial^2 SNR_t}{\partial \beta_2 \partial \beta_1}, \quad SNR_{\beta_2 \beta_2} = \frac{\partial^2 SNR_t}{\partial^2 \beta_2}
\end{eqnarray*}
and the determinant of $2 \times 2$ Hessian matrix as
\begin{equation*}
C(\beta_1, \beta_2) = SNR_{\beta_1 \beta_1} SNR_{\beta_2 \beta_2} - SNR_{\beta_1 \beta_2} SNR_{\beta_2 \beta_1}
\end{equation*}

First, consider the set of points $S_{\beta_1 \beta_2} = \{(\beta_1, \beta_2): (\beta_1, \beta_2) \mbox{ satisfies } \eqref{eqn:b1b2}\}$. For all points in $S_{\beta_1 \beta_2}$ we can prove that
\begin{eqnarray*}
SNR_{\beta_1 \beta_1} &>& 0\\
C(\beta_1, \beta_2) &=& 0
\end{eqnarray*}
Therefore, the second partial derivative test to determine if the set of stationary points $S_{\beta_1 \beta_2}$ of the objective function are local minimum, maximum, or saddle points fails. However, we can establish that for every $(\beta_1, \beta_2) \in S_{\beta_1 \beta_2}$, the following holds
\begin{eqnarray}
\pd{SNR_t}{\beta_1}\bigg|_{(\beta_1+\delta_1, \beta_2+\delta_2)} &<& 0, \quad \pd{SNR_t}{\beta_2}\bigg|_{(\beta_1+\delta_1, \beta_2+\delta_2)} < 0, \quad h_{s1} h_{1t} \delta_1 + h_{s2} h_{2t} \delta_2 < 0, \label{eqn:ineq1}\\
\pd{SNR_t}{\beta_1}\bigg|_{(\beta_1+\delta_1, \beta_2+\delta_2)} &>& 0, \quad \pd{SNR_t}{\beta_2}\bigg|_{(\beta_1+\delta_1, \beta_2+\delta_2)} > 0, \quad h_{s1} h_{1t} \delta_1 + h_{s2} h_{2t} \delta_2 > 0, \label{eqn:ineq2}\\
C(\beta_1+\delta_1, \beta_2+\delta_2) &>& 0, \quad h_{s1} h_{1t} \delta_1 + h_{s2} h_{2t} \delta_2 < 0, \label{eqn:ineq3}\\
C(\beta_1+\delta_1, \beta_2+\delta_2) &>& 0, \quad h_{s1} h_{1t} \delta_1 + h_{s2} h_{2t} \delta_2 > 0, \label{eqn:ineq4}
\end{eqnarray}
for all $(\delta_1, \delta_2) \rightarrow 0$. In other words, \eqref{eqn:ineq1} and \eqref{eqn:ineq2} imply that the slope of the function changes sign at $h_{s1} h_{1t} \beta_1 + h_{s2} h_{2t} \beta_2 = 0$, and \eqref{eqn:ineq3} and \eqref{eqn:ineq4} imply that the convexity of the function, however, does not change at $h_{s1} h_{1t} \beta_1 + h_{s2} h_{2t} \beta_2 = 0$. Therefore, together these imply that \eqref{eqn:b1b2} leads to a local minimum of the objective function.

Next, consider the set of points defined by \eqref{eqn:b1itob2} and \eqref{eqn:b2itob1}. For all such points we can prove that
\begin{eqnarray*}
SNR_{\beta_1 \beta_1} &<& 0\\
C(\beta_1, \beta_2) &>& 0
\end{eqnarray*}
Therefore, from the second partial derivative test the objective function attains it local maximum at the set of point characterize by \eqref{eqn:b1itob2} and \eqref{eqn:b2itob1} above. However, no real solution of the simultaneous system of equation in \eqref{eqn:b1itob2} and \eqref{eqn:b2itob1} exists. In other words, no solution of \eqref{eqn:diamondProb} exists where both relay nodes are transmitting strictly below their respective transmit power constraints.

Above discussion implies that all points satisfying \eqref{eqn:b1b2} lead to the global minimum of the objective function in \eqref{eqn:diamondProb} and the global maximum of the objective function occurs either at planes defined by $\beta_1 = \beta_{1,max}$ or $\beta_2 = \beta_{2,max}$ or at the corner-point $(\beta_{1,max}, \beta_{2,max})$. Therefore all choices of the parameters $(\{h_e, e \in E\}, P_s, P_1, P_2)$ that result in one of the constraints $\beta_1^2 < \beta_{1,max}^2$ at $\beta_2 = \beta_{2,max}$ plane and $\beta_2^2 < \beta_{2,max}^2$ at $\beta_1 = \beta_{1,max}$ being satisfied lead to a whole class of scenarios where global optimum solutions are achieved when the transmit power of one relay node is less than the corresponding maximum, thus contradicting assumption A. For example $(h_{s1} = h_{1t} = h_{2t} = 1, h_{s2} = 0.1, P_s = P_1 = P_2 = 10)$ leads to the optimal solution $(\beta_1 = 0.995, \beta_2 = 0.225)$ whereas $(\beta_{1,max} = 0.995, \beta_{2,max} = 7.07)$, as we show in \cite{111agnihotriJaggiChen}.  {\hspace*{\fill}~\IEEEQEDclosed\par}

\section{Computing $\bm{\beta}_{opt}$ layer-by-layer}
\label{sec:betaComputation}
In this section we prove that in an end-to-end rate optimal network-wide scaling vector $\bm{\beta}_{opt}$ in \eqref{eqn:eqProb}, the component scaling factors corresponding to the relay nodes in the layer $l, 1 \le l \le L$, maximize the sum-rate of the nodes in the layer $l+1$. However, before discussing this result formally, we motivate it by computing the maximum rate of information transfer over a linear amplify-and-forward relay network.

\subsection{Linear AF Networks}
\label{subsec:linNets}
\begin{figure}[!t]
\centering
\includegraphics[width=4.0in]{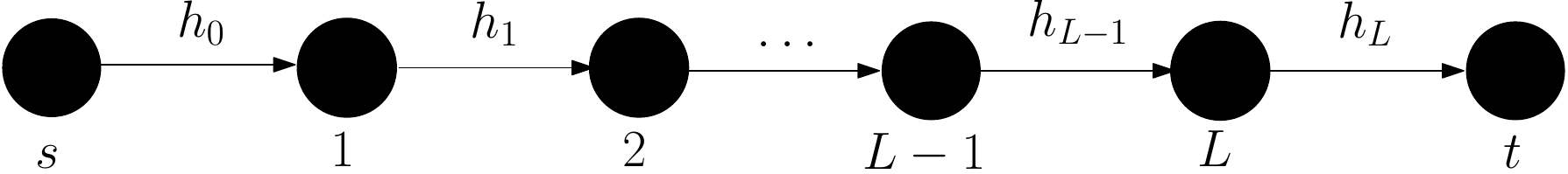}
\caption{A linear amplify-and-forward relay network of $L+2$ layers, with source $s$ in layer `$0$', destination $t$ in layer `$L+1$', and $L$ relay nodes between them.}
\label{fig:linNet}
\end{figure}

We consider a linear amplify-and-forward network of $L$ relay nodes between source $s$ and destination $t$, as shown in Figure~\ref{fig:linNet}.

Let a feasible scaling vector $\bm{\beta}=(\beta_1, \ldots, \beta_L)$ be such that the output of each relay node satisfies the corresponding transmit power constraint \eqref{eqn:pwrConstraint}. Then the maximum scaling factor for the $l^\textrm{th}, 1 \le l \le L$, relay is (from \eqref{eqn:AFdef}):
\begin{equation}
\label{eqn:betaMax}
\beta_{l,max}^2 = \frac{P_l}{P_s (h_0 \prod_{i=1}^{l-1} \beta_i h_i)^2 + \sigma^2(1 + \sum_{i=1}^{l-1}(\prod_{j=i}^{l-1} \beta_j h_j)^2)}
\end{equation}

In a linear AF network, both the source signal and the noise introduced at each intermediate relay node can reach the destination along only one path. Therefore using \eqref{eqn:modChnlParams}, \eqref{eqn:modChnlParams2}, \eqref{eqn:chnlmod}, and \eqref{eqn:snr}, for a given scaling vector $\bm{\beta}$, the received SNR at destination $t$ or any relay node can be written as
\begin{equation}
\label{eqn:linNetnodeSNR}
SNR_l = \frac{P_s}{\sigma^2} \frac{(h_0 \prod_{i=1}^{l-1} \beta_i h_i)^2}{1 + \sum_{i=1}^{l-1}(\prod_{j=i}^{l-1} \beta_j h_j)^2}, 1 \le l \le L+1
\end{equation}

\begin{pavikl}
\label{lemma:betaN-1optimality}
The value of $\beta_{L-1}$ that maximizes $SNR_L$ also maximizes $SNR_t$.
\end{pavikl}
\begin{IEEEproof}
The proof involves three steps.

\textit{Step 1:} Consider the partial derivative of $SNR_t$ with respect to $\beta_L$:
\begin{equation*}
\pd{SNR_t}{\beta_L} = 2 \frac{P_s h_0^2}{\sigma^2} \frac{(\prod_{i=1}^{L-1} \beta_i h_i)^2 \beta_L h_L^2}{(1 + \sum_{i=1}^L(\prod_{j=i}^L \beta_j h_j)^2)^2} 
\end{equation*}
This implies that for a given $(\beta_1, \ldots, \beta_{L-1})$, $SNR_t$ increases with $\beta_L$. However, as the maximum value that $\beta_L$ can take is $\beta_{L,max}$, so $SNR_t$ attains it maximum value at $\beta_{L,max}$.

\textit{Step 2:} Using \eqref{eqn:betaMax} we can express $SNR_t$ only in terms of $(\beta_1, \ldots, \beta_{L-1})$ as $SNR_t(\beta_1, \ldots, \beta_{L-1})$ given below as
\begin{equation*}
SNR_t(\beta_1, \ldots, \beta_{L-1}) = \frac{\frac{P_s h_0^2 P_L h_L^2}{\sigma^2}}{P_s h_0^2 + \frac{\sigma^2 + P_L h_L^2}{(\prod_{i=1}^{L-1} \beta_i h_i)^2}(1 + \sum_{i=1}^{L-1}(\prod_{j=i}^{L-1} \beta_j h_j)^2)}
\end{equation*}

\textit{Step 3:} Compute the partial derivative of $SNR_t(\beta_1, \ldots, \beta_{L-1})$ with respect to $\beta_{L-1}$ as
\begin{equation}
\label{eqn:destSNRdwrtbetaL-1}
\pd{SNR_t(\beta_1, \ldots, \beta_{L-1})}{\beta_{L-1}} = \frac{\frac{P_s h_0^2 P_L h_L^2}{\beta_{L-1}}(1+\frac{P_L h_L^2}{\sigma^2})}{\big[P_s h_0^2 + \frac{\sigma^2 + P_L h_L^2}{(\prod_{i=1}^{L-1} \beta_i h_i)^2}(1 + \sum_{i=1}^{L-1}(\prod_{j=i}^{L-1} \beta_j h_j)^2)\big]^2}
\end{equation}

Further, from \eqref{eqn:linNetnodeSNR} the partial derivative of $SNR_{L}$ with respect to $\beta_{L-1}$ evaluates to
\begin{equation}
\label{eqn:NSNRdwrtbetaL-1}
\pd{SNR_L}{\beta_{L-1}} = 2 \frac{P_s h_0^2}{\sigma^2} \frac{(\prod_{i=1}^{L-2} \beta_i h_i)^2 \beta_{L-1} h_{L-1}^2}{(1 + \sum_{i=1}^{L-1}(\prod_{j=i}^{L-1} \beta_j h_j)^2)^2}
\end{equation}

It follows from \eqref{eqn:destSNRdwrtbetaL-1} and \eqref{eqn:NSNRdwrtbetaL-1} that $SNR_t(\beta_1, \ldots, \beta_{L-1})$ and $SNR_L$ are increasing functions of $\beta_{L-1}$. Therefore both attain their respective maximum at $\beta_{L-1, max}$, the maximum value of $\beta_{L-1}$. In other words, the value of $\beta_{L-1}$ that maximizes $SNR_L$ also maximizes $SNR_t$.
\end{IEEEproof}

Following the same sequence of steps as in the proof of above lemma with $SNR_t$ and $SNR_L$ replaced by $SNR_L$ and $SNR_{L-1}$, respectively, we can also prove that the same value of $\beta_{L-2}$ (specifically $\beta_{L-2,max}$) maximizes both, $SNR_L$ and $SNR_{L-1}$. This along with Lemma~\ref{lemma:betaN-1optimality} that allows us to express both, $SNR_L$ and $SNR_t$ as functions of $(\beta_1, \ldots, \beta_{L-2})$, proves that the same value of $\beta_{L-2,max}$ maximizes $SNR_{L-1}, SNR_L$ and $SNR_t$. Furthermore carrying out this reasoning recursively allows us to express $SNR_i, 2 \le i \le L+1$, only in terms of $\beta_1$ and to prove that the same value of $\beta_1$ (specifically $\beta_{1,max}$) maximizes all of them. We summarize this in the following proposition.

\begin{pavikp}
\label{prop:optBetaLinNet}
For a linear AF network, the network-wide scaling vector $\bm{\beta}_{opt}=(\beta_{1}^{opt}, \ldots, \beta_{L}^{opt})$ that solves \eqref{eqn:eqProb} can be computed recursively as
\begin{equation*}
\beta_{i}^{opt} = \argmax_{\beta_{i}^2 \le \beta_{i,max}^2} SNR_{i+1}(\beta_{1}^{opt}, \ldots, \beta_{i-1}^{opt}, \beta_{i}), 1 \le i \le L+1
\end{equation*}
\end{pavikp}

\begin{pavikc}
For a linear AF network with $P_s = P_1 = \ldots = P_L = P$ and $h_0 = h_1 = \ldots = h_L = h$, the maximum achievable information rate $R = {\cal O}(1/L)$.
\end{pavikc}
\begin{IEEEproof}
Using Proposition~\ref{prop:optBetaLinNet}, we can show that
\begin{equation*}
(\beta_{i}^{opt})^2 = \beta_{i,max}^2 = \beta^2 =  \frac{P}{h^2 P + \sigma^2}, 1 \le i \le L
\end{equation*}
Therefore from \eqref{eqn:linNetnodeSNR}, we have
\begin{equation*}
SNR_{t,max} = \bigg(\frac{h^2 P}{\sigma^2}\bigg)^2 \frac{1 - (\beta h)^2}{1 - (\beta h)^{2L+2}} (\beta h)^{2L}
\end{equation*}
This implies that the maximum achievable ANC rate in this case, $R_{ANC} = \frac{1}{2}\log(1 + SNR_t)$ varies asymptotically with $L$ as $R \le \frac{1}{2L}\frac{(h^2 P/\sigma^2)^2}{1 + h^2 P/\sigma^2}$.
\end{IEEEproof}

\subsection{General Layered Networks}
\label{subsec:genLnets}
We now discuss our result for general layered networks (any number of layers, any number of nodes in each layer, and any connectivity matrix between the nodes in adjacent layers) in general SNR regime.

\begin{pavikl}
\label{lemma:optBetaGenNet}
Consider a layered relay network of $L+2$ layers, with source $s$ in layer `$0$', destination $t$ in layer `$L+1$', and $L$ layers of relay nodes between them. The $l^\textrm{th}$ layer contains $n_l$ nodes, $n_0 = n_{L+1} = 1$. A network-wide scaling vector $\bm{\beta}_{opt}=(\bm{\beta}_{1}^{opt}, \ldots, \bm{\beta}_{L}^{opt})$ that solves \eqref{eqn:eqProb} for this network, can be computed recursively for $1 \le l \le L$ as
\begin{equation*}
\bm{\beta}_{l}^{opt} = \argmax_{\bm{\beta}_{l}^2 \le \bm{\beta}_{l,max}^2} \prod_{j=1}^{n_{l+1}}(1+SNR_{l+1, j}(\bm{\beta}_{1}^{opt}, \ldots, \bm{\beta}_{l-1}^{opt}, \bm{\beta}_{l})),
\end{equation*}
where $\bm{\beta}_{l}^{opt}$ is the subvector of optimal scaling factors for the nodes in the $l^\textrm{th}$ layer, $\bm{\beta}_{l}^{opt} = (\beta_{l 1}^{opt}, \ldots, \beta_{l n_l}^{opt})$ and constraints $\bm{\beta}_{l}^2 \le \bm{\beta}_{l,max}^2$ are component-wise $\beta_{l j}^2 \le \beta_{l j,max}^2, 1 \le j \le n_l$.
\end{pavikl}

\textit{Remark 1:} Lemma~\ref{lemma:optBetaGenNet}, in other words, states that the subvector of the optimal scaling vector $\bm{\beta}_{opt}$ corresponding to the scaling factors of the nodes in the $l^\textrm{th}$ layer, is one that maximizes the product $\prod_{j=1}^{n_{l+1}}(1+SNR_{l+1, j})$ over the $n_{l+1}$ nodes in the next $l+1^\textrm{st}$ layer. Now observe that $\log\prod_{j=1}^{n_{l+1}}(1+SNR_{l+1, j})$ corresponds to $\sum_{j=1}^{n_{l+1}} R_{l+1, j}$, the sum of information rates to the nodes in the $l+1^\textrm{st}$ layer. Therefore an interpretation of Lemma~\ref{lemma:optBetaGenNet} is: if starting with the first layer, the scaling factors for the nodes in each successive layer are chosen such that the sum-rate of the nodes in the next layer is maximized, then such a choice also leads to a globally optimal solution of the problem \eqref{eqn:eqProb}.

\textit{Remark 2:} The problem \eqref{eqn:eqProb} is a hard optimization problem in $\sum_l n_l$ variables as noted in Section~\ref{sec:computingANCperf}. However, Lemma~\ref{lemma:optBetaGenNet} leads to a decomposition of this problem into a cascade of $L$ such subproblems, where the $l^\textrm{th}$ subproblem involves $n_l$ variables. This results in exponential reduction in search space required to solve \eqref{eqn:eqProb} in general layered networks. For example, in a layered network with $L$ relay layers, each with $N$ nodes, our result allows us to obtain the solution of a hard problem with $L \cdot N$ variables by solving $L$ cascaded subproblems, each with $N$ variables.

\textit{Remark 3:} Though Lemma~\ref{lemma:optBetaGenNet} leads to significant reduction in the computational effort required to solve the problem \eqref{eqn:eqProb}, it does not alter the worst-case computational complexity of the problem \eqref{eqn:eqProb}. To see this consider the subproblem of computing a set of optimal scaling factors of the nodes in the $l^\textrm{th}$ layer that maximizes the product $\prod_{j=1}^{n_{l+1}}(1+SNR_{l+1, j})$ over the nodes in the $l+1^\textrm{st}$ layer. The objective function $\prod_{j=1}^{n_{l+1}}(1+SNR_{l+1, j})$ of this subproblem itself is a ratio of posynomials in $n_l$ variables, in general. Therefore, following the same argument as in Section~\ref{sec:computingANCperf}, we conclude that each of these subproblems itself is computationally intractable.

\begin{IEEEproof}
For the ease of presentation, we discuss the proof for a class of layered networks where channel gains along all links between the nodes in two adjacent layers are equal, as in Figure~\ref{fig:layrdNet4proof}. We call such layered networks as \textit{``Equal Channel Gains between Adjacent Layers (ECGAL)''} networks. In particular, we discuss the proof for the ECGAL network shown in Figure~\ref{fig:layrdNet4proof}. We assume that all nodes have the same transmit power constraint $EX^2 \le P$. Consider three adjacent layers $k-1, k$, and $k+1$.

\begin{figure}[!t]
\centering
\includegraphics[width=3.5in]{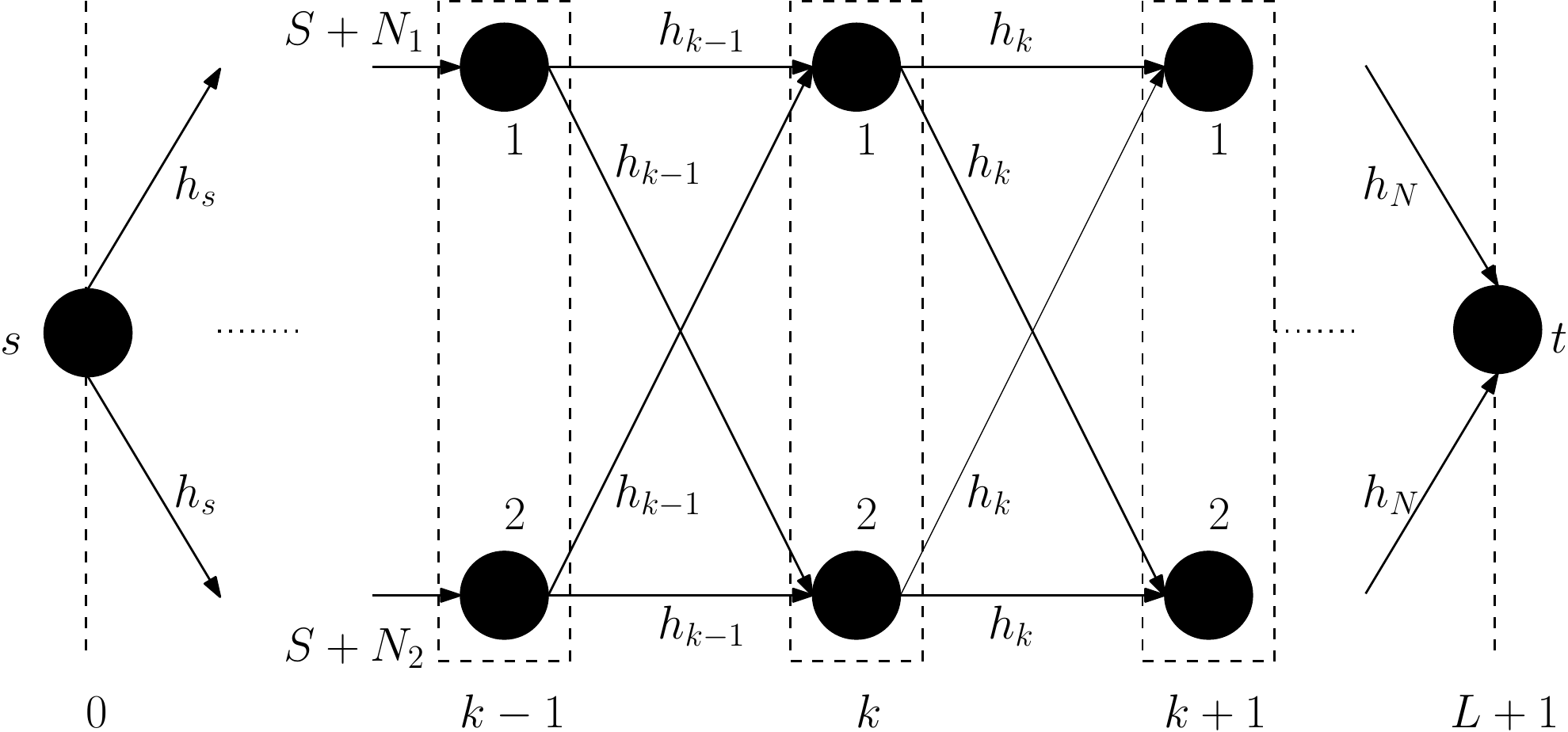}
\caption{An ECGAL network of $L+2$ layers, with source $s$ in layer `$0$', destination $t$ in layer `$L+1$', and $L$ layers consisting of two relay nodes each between them. The channel gains along all links between two adjacent layers are equal.}
\label{fig:layrdNet4proof}
\end{figure}

\textbf{\textit{Claim:}} The scaling factors for the nodes in layer $k-1$ that maximize $\prod_{j=1}^{2}(1+SNR_{k, j})$ also maximize $\prod_{j=1}^{2}(1+SNR_{k+1, j})$ and vice-versa.

\textit{Proof:} Let the source signal components\footnote{Given the symmetry of the ECGAL network, source signals at the input of the nodes in every layer are identical.} of the input at the two nodes in the layer $k-1$ be denoted as $S$, with $var(S) = S^2$. Let the noise components at the two nodes be denoted as $N_1$ and $N_2$, respectively, with $var(N_1) = var(N_2) = N^2$.

The SNRs at the nodes in layers $k$ and $k+1$ are given as:
\begin{eqnarray*}
SNR_{k,1} &=& SNR_{k,2} = \frac{\alpha^2}{\gamma^2} \\
SNR_{k+1,1} &=& SNR_{k+1,2} = \frac{\alpha^2 h_k^2 (\beta_3 + \beta_4)^2}{\sigma^2 + \gamma^2 h_k^2 (\beta_3^2 + \beta_4^2)},
\end{eqnarray*}
with $\alpha^2 = S^2 h_{k-1}^2 (\beta_1 + \beta_2)^2$ and $\gamma^2 = \sigma^2 + h_{k-1}^2 N^2 (\beta_1^2 + \beta_2^2)$.

Define
\begin{eqnarray*}
SNR_{k} &=& \prod_{j \in \{1,2\}}(1+SNR_{k,j}) = (1+SNR_{k,1})^2 \\
SNR_{k+1} &=& \prod_{j \in \{1,2\}}(1+SNR_{k+1,j}) = (1+SNR_{k+1,1})^2
\end{eqnarray*}

First let us consider the problem
\begin{equation}
\label{eqn:SNRL3wb3b4}
\max_{\beta_{k,j}^2 \le \beta_{k,max}^2} SNR_{k+1},
\end{equation}
where $\beta_{k,max}^2 = \frac{P}{\alpha^2 + \gamma^2}, j \in \{1,2\}$.

Evaluating the partial derivatives of $SNR_{k+1}$ with respect to $\beta_{k,1}$ and $\beta_{k,2}$ and setting those equal to zero, we get the following three conditions for global extrema:
\begin{eqnarray}
\beta_{k,1} &=& -\beta_{k,2} \label{eqn:b3b4}\\
\beta_{k,1} &=& \frac{\sigma^2}{\gamma^2 h_k^2 \beta_{k,2}} + \beta_{k,2} \label{eqn:b3itob4}\\ 
\beta_{k,2} &=& \frac{\sigma^2}{\gamma^2 h_k^2 \beta_{k,1}} + \beta_{k,1} \label{eqn:b4itob3}
\end{eqnarray}

It can be proved that \eqref{eqn:b3b4} leads to the global minimum of the objective function in \eqref{eqn:SNRL3wb3b4} and no solution of \eqref{eqn:b3itob4} and \eqref{eqn:b4itob3} exists on or within the boundaries of feasible region defined by the constraints $0 \le \beta_{k,1}^2 \le \beta_{k,max}^2$ and $0 \le \beta_{k,2}^2 \le \beta_{k,max}^2$. Therefore only solution of \eqref{eqn:SNRL3wb3b4} is attained at the corner $(\beta_{k,max}^2, \beta_{k,max}^2)$ of the feasible region.

Substituting the above solution of \eqref{eqn:SNRL3wb3b4} in the expression for $SNR_{k+1}$ above, allows us to express it in terms of $\beta_{k-1,1}$ and $\beta_{k-1,2}$ as $SNR_{k+1}^{\prime}$. Now consider solutions of the following two problems.
\begin{eqnarray}
\max_{\beta_{k-1,j}^2 \le \beta_{k-1,max}^2} SNR_{k+1}^{\prime}, \label{eqn:SNRL3wb1b2} \\
\max_{\beta_{k-1,j}^2 \le \beta_{k-1,max}^2} SNR_{k}, \label{eqn:SNRL2wb1b2},
\end{eqnarray}
where $\beta_{k-1,max}^2 = \frac{P}{S^2 + N^2}, j \in \{1,2\}$.

Computing the partial derivatives of $SNR_{k+1}$ and $SNR_{k}$ with respect to $\beta_{k-1,1}$ and $\beta_{k-1,2}$ and equating those to zero, we get the following three conditions for their respective global extrema:
\begin{eqnarray}
\beta_{k-1,1} &=& -\beta_{k-1,2} \label{eqn:proofb1b2}\\
\beta_{k-1,1} &=& \frac{\sigma^2}{h_{k-1}^2 N^2 \beta_{k-1,2}} + \beta_{k-1,2} \label{eqn:proofb1itob2}\\
\beta_{k-1,2} &=& \frac{\sigma^2}{h_{k-1}^2 N^2 \beta_{k-1,1}} + \beta_{k-1,1} \label{eqn:proofb2itob1}
\end{eqnarray}

It can be proved that \eqref{eqn:proofb1b2} leads to the global minima of the problems \eqref{eqn:SNRL3wb1b2} and \eqref{eqn:SNRL2wb1b2} and no solution of \eqref{eqn:proofb1itob2} and \eqref{eqn:proofb2itob1} exists on or within the boundaries of feasible region defined by the constraints $0 \le \beta_{k-1,1}^2 \le \beta_{k-1,max}^2$ and $0 \le \beta_{k-1,2}^2 \le \beta_{k-1,max}^2$. Therefore only solution of both the problems \eqref{eqn:SNRL3wb1b2} and \eqref{eqn:SNRL2wb1b2} is attained at the corner $(\beta_{k-1,max}^2, \beta_{k-1,max}^2)$ of the feasible region. Thus proving our claim. {\hspace*{\fill}~\IEEEQEDopen\par}

Carrying out the above procedure in the proof of our claim recursively for all $k, 1 \le k \le L$ layers, proves the theorem for the ECGAL networks we consider here.
\end{IEEEproof}

\section{Illustration}
\label{sec:exa}
In the following, we illustrate the usefulness of Lemma~\ref{lemma:optBetaGenNet} by computing the maximum achievable ANC rate in a network scenario without any \textit{a priori} assumption on input signal scaling factors and the received SNRs, as in \cite{110maricGoldsmithMedard, 111liuCai}.

\textbf{\textit{Example 3:}} Consider the ECGAL network of Figure~\ref{fig:layrdNet4proof} with $L$ layers of relay nodes between source $s$ and destination $t$, and $N$ nodes in each layer. We assume that the channels gains along all links are equal and denoted as $h$. We also assume that all nodes have the same transmit power constraint $EX^2 \le P$. The SNR at destination $t$ for this network is
\begin{equation*}
SNR_t = \frac{h^2 P}{\sigma^2} \frac{h^{2L}(\sum_{p \in \mathcal{P}_{s,t}} \prod_{p:(i_1, \ldots, i_L)} \beta_{i_1} \ldots \beta_{i_L})^2}{1 + \sum_{l=1}^L \sum_{j=1}^N h^{2(L-l+1)}\big[\beta_{lj} \sum_{p \in \mathcal{P}_{lj,t}} \prod_{p:(i_{l+1}, \ldots, i_L)} \beta_{i_{l+1}} \ldots \beta_{i_L}\big]^2},
\end{equation*}
where $\mathcal{P}_s$ denotes the set of all $N^L$ paths from source $s$ to destination $t$ over $L$ layers, each with $N$ relay nodes; $\mathcal{P}_{lj}$ denotes the set of all $N^{L-l}$ paths from the $j^\textrm{th}$ node in the $l^\textrm{th}$ layer to destination $t$, and $(i_k, \ldots, i_L)$ denotes the set of indices of the nodes belonging to the path under consideration. Therefore, the problem of maximum rate achievable with analog network coding for this network can be formulated as (using \eqref{eqn:eqProb})
\begin{equation}
\label{eqn:ecgalExa}
\argmax_{0 \le \bm{\beta}^2 \le \bm{\beta}_{max}^2} SNR_t,
\end{equation}
where $\bm{\beta}=(\beta_{lj})_{1 \le l \le L, 1 \le j \le n_l}$ and constraints $\bm{\beta}^2 \le \bm{\beta}_{max}^2$ are component-wise $\beta_{l j}^2 \le \beta_{l j,max}^2$, and
\begin{equation*}
\beta_{lj,max}^2 = \frac{P_{lj}}{P(h^L \sum\limits_{p \in \mathcal{P}_{s,lj}} \prod\limits_{p:(i_1, \ldots, i_{l-1})} \beta_{i_1} \ldots \beta_{i_{l-1}})^2 + \sigma^2 (1 + \sum_{k=1}^{l-1} \sum_{m=1}^N h^{2(l-k)}\big[\beta_{km} \sum\limits_{p \in \mathcal{P}_{km,lj}} \prod\limits_{p:(i_{1}, \ldots, i_{l-1})} \beta_{i_1} \ldots \beta_{i_{l-1}}\big]^2)},
\end{equation*}
where $\mathcal{P}_{s,lj}$ denotes the set of all $N^{l-1}$ paths from source $s$ to the $j^\textrm{th}$ node in the $l^\textrm{th}$ layer over $l-1$ intervening layers, each with $N$ relay nodes; and $\mathcal{P}_{km,lj}$ denotes the set of all $N^{l-k-1}$ paths from the $m^\textrm{th}$ node in the $k^\textrm{th}$ layer to the $j^\textrm{th}$ node in the $l^\textrm{th}$ layer over $l-k-1$ intervening layers.

From the symmetry of the network, it follows that $\beta_{li, max}^2 = \beta_{l,max}^2$, $1 \le l \le L, 1 \le i \le N$, where
\begin{equation*}
\beta_{l,max}^2 = \frac{P/\sigma^2}{\big[h \prod\limits_{i=1}^{l-1}(N \beta_i h)\big]^2 \frac{P_s}{\sigma^2} + N \sum\limits_{i=1}^{l-1} (\beta_i h  \prod\limits_{j=i+1}^{l-1}(N \beta_j h))^2 + 1}
\end{equation*}
Using Lemma~\ref{lemma:optBetaGenNet}, we can solve problem \eqref{eqn:eqProb} for this network. The solution $\bm{\beta}_{opt}^N$ is such that all relays in a layer use the same scaling factor and it is equal to the maximum value of the scaling factor for the nodes in the layer, {\it i.e.} $\beta_{li}^2 = \beta_{l,max}^2$, $1 \le l \le L$, and $1 \le i \le N$. The corresponding $SNR_t$ is:
\begin{equation}
\label{eqn:ecgalExa}
SNR_{t,opt} = \frac{h^2 P_s}{\sigma^2} \frac{(Nh)^{2L} \prod_{l=1}^L \beta_l^2}{1 + N h^2 \sum_{l=1}^L (Nh)^{2(L-l)} \prod_{i=l}^L \beta_i^2}
\end{equation}
and the maximum achievable ANC rate in this scenario is $R_{ANC} = \frac{1}{2}\log(1 + SNR_{t,opt})$. In the following, we discuss the computation of $R_{ANC}$ in two scenarios.

\textit{Case 1:} Let $P_s \rightarrow 0$, then for the leading order in $N$:
\begin{equation*}
SNR_{t,opt} = \frac{N^2 P_s}{\sigma^2} \frac{N h^2 P}{\sigma^2} \frac{1}{1+L/N}
\end{equation*}
The received SNR at the $l^\textrm{th}$ layer varies with the number of preceding layers as $SNR \sim (1+\frac{l-1}{N})^{-1}$. Therefore for any fixed $\delta$ as in \cite{110maricGoldsmithMedard, 111liuCai}, an arbitrarily large number of layers may violate the high-SNR regime condition $\min_{k \in l} P_{R,k} \ge 1/\delta, l = 1, \ldots, L$ as $L$ grows. Thus the approaches in \cite{110maricGoldsmithMedard, 111liuCai} cannot be used to exactly compute $SNR_{t,opt}$ as above or the optimal ANC rate in such networks.

\begin{figure}[!t]
\centering
\includegraphics[width=3.5in]{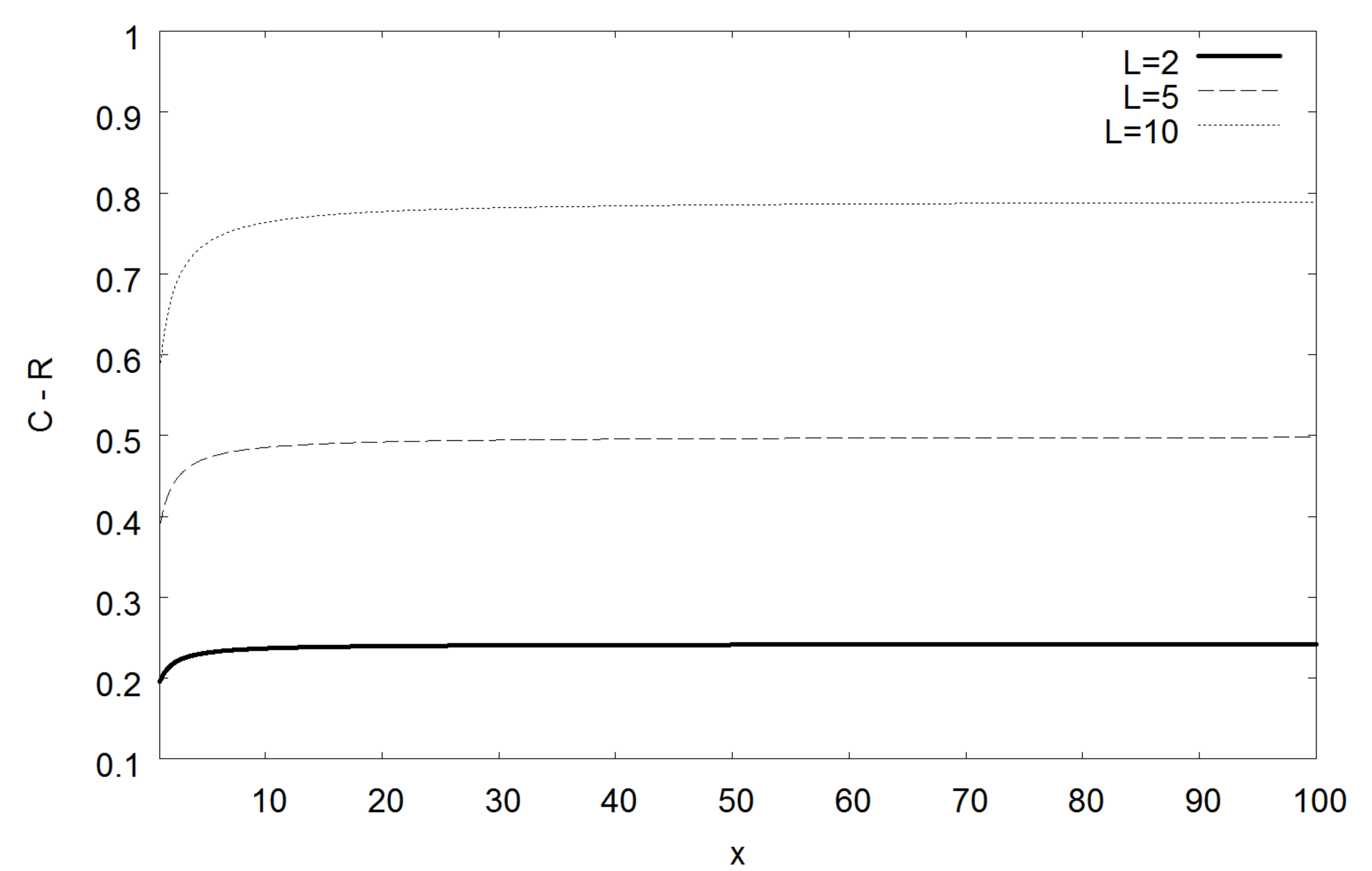}
\caption{For the ECGAL network in Example 2, Case 2: the gap $C-R$ between MAC cut-set bound $C$ and analog network coding rate $R = R_{ANC}$ as parameter $x = \frac{N h^2 P}{\sigma^2}$ increases. The number of nodes in each layer is $N=5$. We observe that for a given number of layers in the network the gap approaches a constant value. As the number of layers in the network increases, the corresponding gap also increases.}
\vspace{-0.2in}
\label{fig:gap2mac}
\end{figure}

\textit{Case 2:} Let $P_s \rightarrow \infty$. In this case, for the leading order in $N$ we have
\begin{equation*}
SNR_{t,opt} = N x \frac{1}{1+L/N}, \quad x = \frac{N h^2 P}{\sigma^2}
\end{equation*}
Therefore, $R_{ANC} = \frac{1}{2}\log(1 + SNR_{t,opt})$ approaches the MAC cut-set bound $C = \frac{1}{2} \log(1+ N x)$ \cite{111agnihotriJaggiChen}, within a constant gap as $x \rightarrow \infty$, as shown in Figure~\ref{fig:gap2mac}.

\section{Conclusion and Future Work}
\label{sec:conclFW}
We consider the problem of maximum rate achievable with analog network coding in general layered networks. Previously, this problem was addressed assuming that the nodes in all but at most one layer in the network are in the high-SNR regime, and each node forwards the received signal at the upper bound of its transmit power constraint. We provide a key result that allows us to exactly compute the maximum ANC rate in a class of layered network without these two assumptions. Further, our result significantly reduces the computational complexity of this problem for general layered networks. We illustrate the significance of our result by computing the maximum ANC rate for one particular layered relay network in a scenario that cannot be addressed using existing approaches. In the future, we plan to extend this work to general wireless networks.

\end{document}